\documentclass[aps,prb,twocolumn,showpacs,10pt]{revtex4-1}
\usepackage{graphicx,amsmath,amssymb,amsfonts,sidecap,color}

\makeatletter
\newcommand*{\rom}[1]{\expandafter\@slowromancap\romannumeral #1@}
\makeatother

\begin{document}

\title{Topological Edge States and Fractional Quantum Hall Effect  from Umklapp Scattering}

\author{Jelena Klinovaja}
\affiliation{Department of Physics, University of Basel,
             Klingelbergstrasse 82, CH-4056 Basel, Switzerland}
\author{Daniel Loss}
\affiliation{Department of Physics, University of Basel,
             Klingelbergstrasse 82, CH-4056 Basel, Switzerland}

\date{\today}
\pacs{71.10.Fd; 05.30.Pr; 71.10.Pm; 73.43.-f }



\begin{abstract}
We study anisotropic lattice strips
in the presence of a magnetic field in the quantum Hall effect regime. At specific magnetic fields, causing resonant Umklapp scattering, the system is gapped in the bulk and supports chiral edge states in close analogy to topological insulators. In electron gases with stripes, these gaps result in plateaus for the Hall conductivity  exactly at the known fillings $n/m$ (both positive integers and $m$ odd) for the integer and fractional quantum Hall effect.
For double strips we find topological phase transitions with phases that support midgap edge states with flat dispersion.
The topological effects predicted here could be tested directly in optical lattices.
\end{abstract}

\maketitle

{\it Introduction.} Condensed matter systems with topological properties have attracted wide attention over the years. \cite{Wilczek, Hasan_RMP, Zhang_RMP,Alicea_2012} E.g., the integer and fractional  quantum Hall effects (IQHE and FQHE) \cite{Klitzing,Tsui_82}  find their origin in the  topology of the system. \cite{QHE_Review_Prange, book_Jain,Hofstadter, Laughlin,Streda,Thouless,Halperin,Laughlin_FQHI,Halperin_crystall,oded_2013,Kane_lines}
Similarly, band insulators with topological properties have become of central interest recently,~\cite{Hasan_RMP,Zhang_RMP,Topological_class_Ludwig}
as well as exotic topological states like fractionally charged fermions \cite{Jackiw_Rebbi,FracCharge_Su,FracCharge_Kivelson,FracCharge_Chamon,CDW, Two_field_Klinovaja_Stano_Loss_2012, Klinovaja_Loss_FF_1D, Frac_graphene_2007, Franz_2009} or Majorana fermions.
\cite{Read_2000,Nayak,fu, Nagaosa_2009, Sato, lutchyn_majorana_wire_2010, oreg_majorana_wire_2010,Klinovaja_CNT, Sticlet_2012,bilayer_MF_2012, MF_nanoribbon}

Here, we study two-dimensional (2D)  strips in magnetic fields, both analytically and numerically, 
modeled by an anisotropic tight-binding lattice. 
We identify a striking mechanism by which
 the magnetic field induces resonant Umklapp scattering (across Brillouin zones) that opens a gap in the bulk spectrum and results in chiral edge states 
in analogy to topological insulators.
 Quite remarkably, the resonant scattering occurs at well-known filling factors for the IQHE \cite{Klitzing} and FQHE \cite{Tsui_82} $\nu=n/m$, where $n,m$  are positive integers and $m$ odd.
We argue below that this mechanism could shed new light on the QHE for 2D electron gases as well, where  the formation e.g. of a periodic structure (energetically favored also by  a Peierls transition) might support the periodic structure needed for the Umklapp scattering.

Finally, we consider a double strip of spinless fermions, or, equivalently, a single strip with spinful fermions.
Here, we find two topological phase transitions accompanied by a closing and reopening of the bulk gap, and, as a result, three distinct phases.  The trivial phase is without edge states. The first topological phase is similar to the one discussed above and carries two propagating chiral  modes at each edge for $\nu=1$. The second topological phase has only one  state at each edge. Quite remarkably, its dispersion is flat throughout the  Brillouin zone, making this phase an attractive playground for studying interaction effects.

\begin{figure}[!bt]
 \includegraphics[width=\columnwidth]{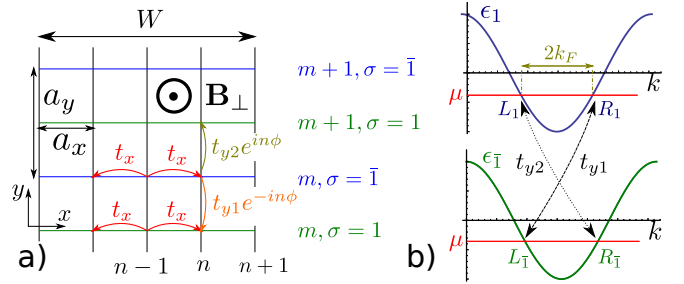}\\
 \caption{(a) Strip: two-dimensional lattice of  width  in $x$-direction, $W$, with  unit cell defined by the lattice constants $a_x$ and $a_y$. 
  The  hopping amplitudes in $y$-direction,  $t_{y1}$ and $t_{y2}$,  carry the phase $\phi$, arising from a perpendicular magnetic field
${\bf B}$, and we assume  $t_x\gg t_{y1},t_{y2}$.
(b) Doubly degenerate spectrum of  $H_x$ [see Eqs. (\ref{1d_x})]
  for the rows with $\sigma=1$ (upper blue) and with $\sigma=\bar{1}$ (lower green). The hoppings  $t_{y1}$ (dashed line) and $t_{y2}$ (dotted line) 
 induce resonant scattering between right ($R_\sigma$) and left ($L_{\bar \sigma}$) movers, 
 which open  gaps at the Fermi wavevectors $\pm k_F$ defined by the chemical potential $\mu$. }
 \label{fig:2d}
\end{figure}

{\it Anisotropic tight-binding model.} We consider a 2D tight-binding model of a strip that is of  width $W$ in  $x$- and  extended in  $y$-direction, see Fig. \ref{fig:2d}a. The unit cell  is composed of two lattice sites ($\sigma=\pm 1$) along $y$ that are distinguished by two  hopping amplitudes, $t_{y1}$ and $t_{y2}$.
Every site is labeled by three indices $n, m$, and $\sigma$, where $n$ ($m$) denotes the position of the unit cell along the $x$- ($y$-) axis. 
The hopping along $x$ is described by
\begin{align}
H_{x} =-t_x \sum_{n, m, \sigma} (c^\dagger_{n+1, m,\sigma} & c_{n,m, \sigma} + h.c.),
\label{1d_x}
\end{align}
where $t_x$ is the hopping amplitude in $x$-direction and $c_{n,m, \sigma} $  the annihilation operator acting on a spinless fermion at
site  $(n,m, \sigma)$, and the sum runs over all sites. 
The hopping along  $y$ is described by
\begin{align}
&H_{y} = \sum_{n, m} (t_{y1} e^{-in\phi}c^\dagger_{n, m,1}  c_{n,m, \bar 1} \nonumber\\
&\hspace{80pt}+ t_{y2}e^{in\phi} c^\dagger_{n, m+1,1}  c_{n,m, \bar 1}+ h.c.),
\label{1d_y}
\end{align}
Without loss of generality, we consider $t_{y2} \geq t_{y1} \geq 0 $. The phase $\phi$ is generated by a uniform magnetic field ${\bf B}$ applied in perpendicular $z$-direction, see Fig. \ref{fig:2d}. 
We choose the corresponding vector potential $\bf A$,
 to be along the $y$-axis, ${\bf A}=(B x) {\bf e}_{y}$, yielding the phase
$\phi = e B  a_x a_y/2\hbar c$.
Here, $a_{x,y}$ are the corresponding lattice constants.

{\it Chiral edge states.}
Taking into account translational invariance of the system in $y$-direction, we introduce the momentum $k_y$ via Fourier transformation, see Appendix \ref{appendix_1}.
The Hamiltonians become diagonal in $k_y$, i.e.,
${ H}_{x} =-t_x \sum_{n, k_y, \sigma} (c^\dagger_{n+1, k_y,\sigma}  c_{n,k_y, \sigma} + h.c.)$, 
and
${ H}_{y} = \sum_{n, k_y} [(t_{y1} e^{-in\phi}+ t_{y2} e^{i(n\phi - k_y a_y )})c^\dagger_{n, k_y,1}  c_{n,k_y, \bar 1}  + h.c.]$.
Thus, the eigenfunctions of ${H}={H}_{x}+{H}_{y}$  factorize as
$e^{i k_y y}\psi_{k_y}(x)$, where we focus now on 
$\psi_{k_y}(x)$ and treat $k_y$ as  parameter.

\begin{figure}[!b]
 \includegraphics[width=0.95\columnwidth]{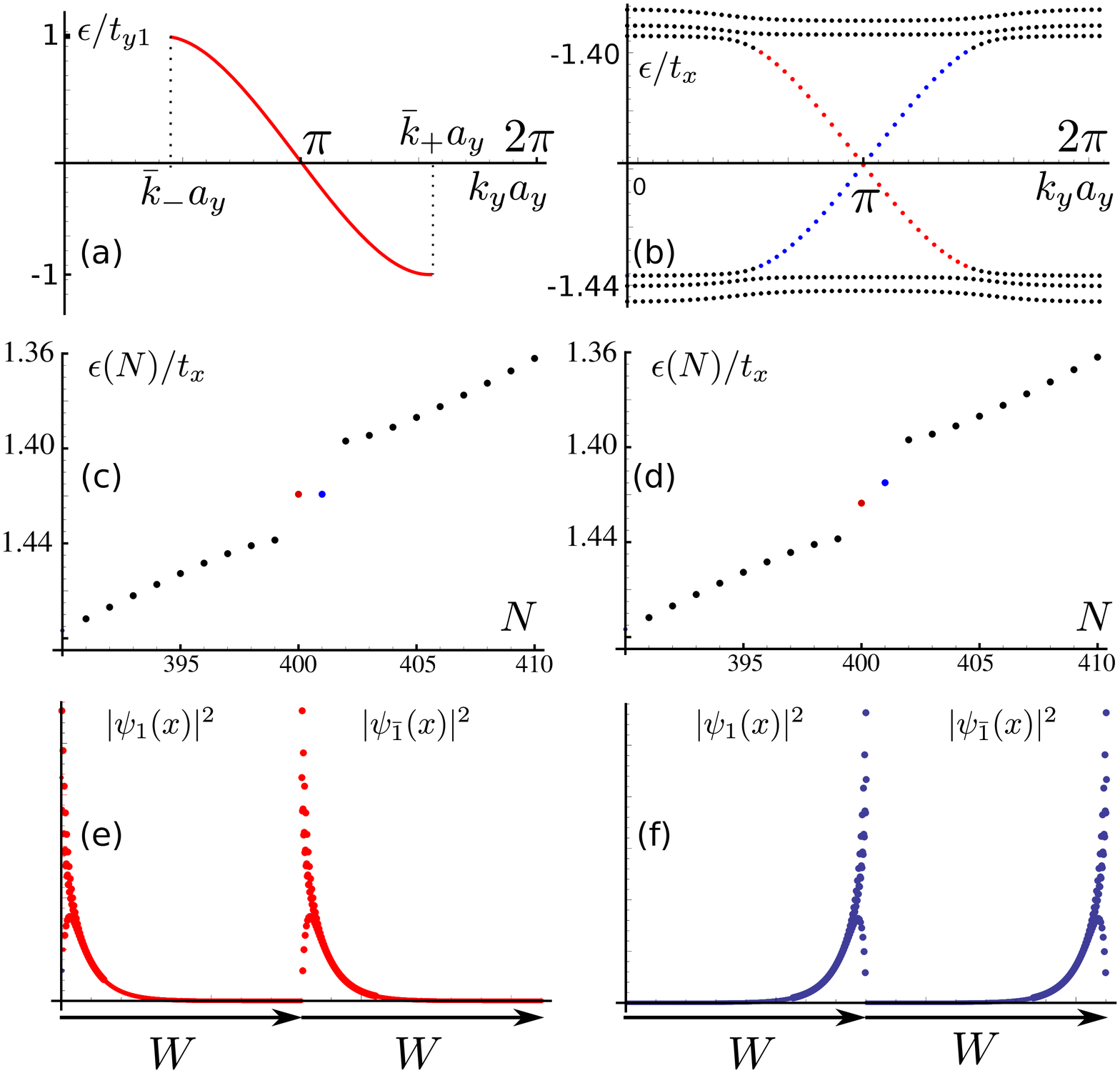}\\
 \caption{Spectrum $E(k_y)$ of  left edge state (red line or  dots) propagating along $y$  for a strip ($t_{y1}/t_x=0.02$, $t_{y2}/t_x=0.1$) of width $W/a_x=801$  
 and with phase $\phi=\pi/2$, obtained (a) analytically [see Eq. (\ref{dispersion})] and (b) numerically [see  Eqs. (\ref{1d_x}) and  (\ref{1d_y})].
For ${\bar k}_-<k_y<{\bar k}_+$, there exists one edge state at each edge. 
The left  (red dots) and the right (blue dots) edge states are chiral  and propagate in opposite $y$-directions. For each  $k_y$ [(c) $k_ya_y=\pi$, (d) $k_ya_y=13\pi/12$] there is one left (red dots)
and one right (blue dots) edge state if  Eq. (\ref{criterion}) is satisfied.  Here, $\epsilon(N)$ corresponds to the $N$th energy level.  The probability density $|\psi_\sigma|^2$  (e) [(f)] of the left [right] localized  state 
decays exponentially in  agreement with the analytical result, Eq. (\ref{wavefunction}) in Appendix ~\ref{appendix_3}.
}
 \label{fig:spectrum}
\end{figure}

Assuming for the moment periodic boundary conditions also in $x$-direction, we introduce a momentum $k_x$, see Appendix \ref{appendix_1}.
Immediately, the well-known spectrum of $H_x$ follows,
$\epsilon_\sigma = - 2 t_x \cos (k_x a_x)$,
which is twofold degenerate in $\sigma$.  
The chemical potential $\mu$ is fixed such that 
 the Fermi wavevector $k_F$ is connected to the phase by $\phi=2k_Fa_x$.
Next, we allow for hopping along $y$ as a small perturbation to the $x$-hopping, {\it i.e.},
$t_x\gg t_{y1}, t_{y2}$, see Fig. \ref{fig:2d}b.
To obtain analytical solutions, it is most convenient to go to the continuum description \cite{Two_field_Klinovaja_Stano_Loss_2012, MF_wavefunction_klinovaja_2012}.
The annihilation operator $\Psi(x)$ close to the Fermi level can be represented in terms of slowly varying right [$R_\sigma (x)$] and left [$L_\sigma (x)$] movers,
$\Psi(x) =\sum_\sigma R_\sigma (x)  e^{ik_F x} + L_\sigma  (x) e^{-ik_F x}$.
The corresponding Hamiltonian density $\mathcal{H}$
can be  rewritten in terms of the Pauli matrices $\tau_i$ ($\sigma_i$), acting on the right-left mover (lattice) subspace (see Appendix \ref{appendix_2}) $\Psi = (R_1, L_1, R_{\bar{1}}, L_{\bar{1}})$, as
\begin{align}
&\mathcal{H}=\hbar \upsilon_F \hat k \tau_3  + \frac{t_{y1}}{2} (\sigma_1 \tau_1+\sigma_2\tau_2)+ \frac{t_{y2}}{2} \Big[(\sigma_1 \tau_1-\sigma_2\tau_2) \nonumber \\
&\hspace{10pt}\times \cos (k_y a_y) + (\sigma_2\tau_1+\sigma_1 \tau_2) \sin (k_y a_y) \Big]. \label{density}
\end{align}
Here,  $\hbar \hat k= -i\hbar \partial_x$ is the momentum operator with eigenvalues $k$ taken from the corresponding Fermi points  $\pm k_F$, and
$\upsilon_F = 2(t_x/\hbar ) a_x \sin(k_F a_x)$ is the Fermi velocity.
The spectrum  with periodic boundary conditions in $x$- and $y$-directions is given by
$\epsilon_{l,\pm} = \pm\sqrt{(\hbar \upsilon_F  k)^2 + t_{yl}^2}$,
where $l=1,2$. This mechanism of opening a gap by oscillatory terms causing resonant scattering between the Fermi points  is similar to a Peierls transition. \cite{Braunecker_Loss_Klin_2009}
Next, we turn to a strip of finite width $W$, see Fig. \ref{fig:2d}. We note that the bulk spectrum $\epsilon_{l,\pm}$ 
is fully gapped, so states localized at the edges can potentially exist. To explore this possibility we consider a semi-infinite nanowire ($x\geq0$) and follow the method developed in Refs. \onlinecite{MF_wavefunction_klinovaja_2012, Two_field_Klinovaja_Stano_Loss_2012}, assuming that the localization length of bound states $\xi$ is much smaller than $W$.
This allows us to impose  vanishing boundary conditions only at $x=0$, $\psi_{k_y}(x)|_{x=0}\equiv(\psi_1, \psi_{\bar 1})|_{x=0}=0$.
This boundary condition is fulfilled only at one energy inside the gap $|E|<t_{y1}$,
\begin{equation}
E(k_y) =  \frac{t_{y1} t_{y2} \sin (k_y a_y)}{\sqrt{t_{y1}^2+t_{y2}^2 - 2 t_{y1} t_{y2} \cos (k_y a_y)}},
\label{dispersion}
\end{equation}
if the following condition is satisfied,
\begin{equation}
t_{y1}>t_{y2} \cos (k_y a_y).
\label{criterion}
\end{equation}
The edge states exist for momenta $k_y \in ({\bar k}_-,{\bar k}_+)$, where $ {\bar k}_\pm a_y =\pi\pm \arcsin (t_{y1}/t_{y2})$. 
An edge state touches a boundary of the gap at ${\bar k}_\pm$ and afterwards disappears in the bulk spectrum of the delocalized states, see Fig. \ref{fig:spectrum}. The only regime in which the edge state exhibits all momenta corresponds to the uniform strip with $t_{y1}=t_{y2}$.
The localization length  $\xi$ is determined by $\xi= \hbar \upsilon_F/\sqrt{t_{y1}^2-E^2}$, with wavefunction given in Appendix \ref{appendix_3}. The edge state  gets delocalized if its energy is close to the boundary of the gap, so that $\xi$ becomes comparable to $W$.
Similarly, we can search for the solution decaying to the right, $x\leq 0$, and obtain Eq. (\ref{dispersion}) with reversed sign, $E(k_y)\to -E(k_y)$.

We have confirmed above results 
by diagonalizing the tight-binding Hamiltonian ${H}$ (in $k_y$-representation)
numerically,
see Fig. \ref{fig:spectrum}. 
The spectrum $E(k_y)$ 
of the edge states localized along $x$ and propagating along $y$  shows that 
at any fixed energy inside the gap there can be only
one edge state at a given edge, see Fig. \ref{fig:spectrum}. Moreover, the edge states are chiral, as can be seen from
the velocity,
$\upsilon = \partial E/\partial k_y$,
which 
is negative (positive)  for the left (right) edge state. 
This means that transport along a given edge  of the strip can occur only in one direction determined by the direction of the $\bf B$-field, see Fig. \ref{fig:2d}. Quite remarkably, the obtained spectrum of edge states is of the same form as for topological insulators \cite{Volkov_Pankratov,Hasan_RMP} with a single  Dirac cone consisting of two crossing non-degenerate subgap modes. Due to the macroscopic separation of opposite edges, $\xi \ll W$, these modes are protected from getting  scattered into each other by impurities, phonons or interaction effects, so that the Dirac cone cannot be eliminated by perturbations that are local and smaller than the gap.  Thus, the edge states are  topologically stable.

{\em Umklapp scattering.}
We note that the system considered here is equivalent to a 2D system in the QHE regime.
The above choice of  magnetic field corresponds to the IQHE with filling factor $\nu=1$, which is in agreement with one chiral mode at each edge. To explore the possibility of inducing quantum Hall physics at other filling factors, we fix the chemical potential $\mu = -\sqrt{2} t_x$, so that the system has local particle-hole symmetry, and change the  B-field. 
Above, the phase $\phi$, generated by the magnetic field, was equal to $\pi/2$ for $k_F =\pi/4a_x$. However, 
this is not the only choice of phase leading to the opening of a gap $\Delta_g$ at the Fermi level. Due to the periodicity of the spectrum, resonant scattering between branches of $\epsilon_\sigma$ occurs also via Umklapp scattering between different Brillouin zones, with a phase
$\phi = \pm \frac{p \pi}{2n}+2\pi q$,
where $q$ is an integer, $n$ a positive integer, and $p$  a positive odd integer with $p < 2n$  and coprime to $n$. As a result, the Fermi level lies in the bulk gap for the  filling factors~\cite{footnote_filling} 
$\nu = n/(4qn\pm p)$, which can be rewritten as $\nu = n/m$, where $m>0$ is an odd integer. 
The size of the gap can be estimated as $\Delta_g \propto t_{yl} (t_{yl}/t_x)^{(n-1)}$ (assuming for simplicity $t_{y1}=t_{y2}$). Finally, we remark that
we checked numerically that the gap $\Delta_g$ never closes for any  finite ratio of $t_x$ and $t_{yl}$ larger or smaller than one.

{\em FQHE in 2D electron gas.} We conjecture that the same mechanism of resonant Umklapp scattering 
can also lead to the 
integer or fractional QHE in 2D electron gases.  At high magnetic fields  interaction effects get strongly enhanced and 
electrons tend to order themselves into periodic structures. 
\cite{Wigner_Girvin, Wigner_Kivelson, Wigner_Jain,book_Jain,anisotropy_CDW,strips_Review,anisotropy_West,QHE_strips}
In particular, we assume the formation of stripes that are aligned along $x$ and periodically repeated in $y$. 
While particles can hop between stripes, they move now continuously inside them  
with quadratic dispersion. 
Thus, the perturbative solutions found above in terms of right- and left-movers still apply.
In addition, we assume that the interaction generates a charge-density wave at wavevector $K$ inside the stripe, providing an effective periodic potential in $x$, which will lead to gaps.
Thus, $K$ becomes the period of the Brillouin zone, and, at $1/4$-filling of the lowest subband, we have $K=8k_F$.~\cite{footnote:B_period}

Again, the $B$-field leads to a gap at $k_F$ only if it results in phases commensurable to $k_F$, i.e. $eB a_y/2\hbar c  = \pm 2k_F \frac{p}{n} + q K$, which is equivalent to 
$\nu =n/(4qn\pm p)\equiv n/m$. 
In this regime, there is an additional energy gain due to a Peierls transition, favoring even more 
a formation of 
periodic structures with gaps. 
 Moreover, from this mean field scenario it follows that the IQHE is more stable against disorder than the FQHE since the latter requires Umklapp scattering through higher Brillouin zones. 
The gap
and the edge states  can be tested in transport experiments. For example, the Hall conductance $\sigma_H$ exhibits plateaus on the classical dependence curve $\sigma_H \propto 1/B$, if the Fermi level lies in the gap.  This can be shown by using the Streda formula, \cite{Streda} $\sigma_H = e c \left( \frac{\partial {\bar n}}{\partial B}\right)_\mu$, where ${\bar n}$ is the bulk particle density which
is uniquely determined by the magnetic field via the relation $\nu e B a_y/2\hbar c=2k_F $ (for this it is crucial that $K$ depends on $k_F$).
If $\mu$ lies in the gap opened by the Umklapp scattering, the change in the density for fixed $\mu$, $d{\bar n}$, due to a change in the magnetic field, $dB$, is given by 
$d{\bar n}  = \frac{d k_F}{\pi  a_y/2} =\nu \frac{e}{h c}  d B$.
Hence, the conductance assumes the  FQHE plateaus,  $\sigma_H = \nu {e^2}/{h c}$, with $\nu=n/m$ and independent of any lattice parameters. The width of the plateaus
is determined by the gap size $\Delta_g \propto t_{yl} (t_{yl}/\mu)^{(n-1)}$.  We note that the FQHE can be mapped back to the IQHE by redefining  the charge $e$ by $e^\star=e/m$ that allows us to keep all scattering events inside the first Brillouin zone. Finally, the distance between stripes can be estimated as
${a_y}/{2} = {k_F}/{\pi \bar{n}}$.

\begin{figure}[!b]
 \includegraphics[width=0.6\columnwidth]{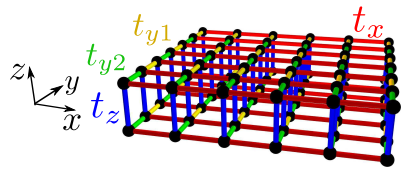}\\
 \caption{Double strip for spinless fermions. The intra-strip couplings are the same as in Fig. \ref{fig:2d}. The inter-strip coupling along $z$ is described by the hopping amplitude $t_z$.}
 \label{fig:3d}
\end{figure}

\begin{figure*}[!th]
 \includegraphics[width=1.4\columnwidth]{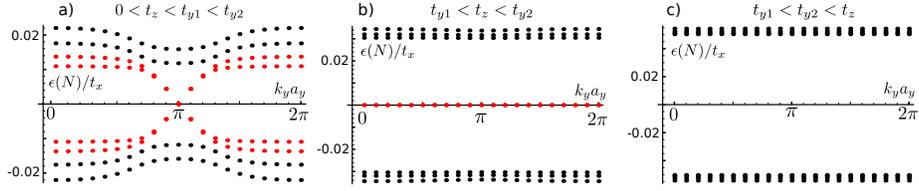}\\
 \caption{Spectrum of a double strip 
 obtained by  numerical diagonalization of the tight-binding Hamiltonian $H_2=H_{x}+H_{y}+H_{z}$ for the same parameters as in Fig. \ref{fig:spectrum}.
 (a) If $|t_z|<t_{y1}$ [$t_z/t_x=0.01$],  there are four edge states at any energy within the gap; two of them localized at the left  and two  at the right edge. (b) If $t_{y1}<|t_z|<t_{y2}$ [$t_z/t_x=0.05$], there is one zero-energy edge state (i.e. with flat dispersion) at each edge. (c) If $|t_z|>t_{y2}$ [$t_z/t_x=1.5$], there are no edge states in the gap.}
 \label{fig:states_3_top}
\end{figure*}

{\it Double strip.}
Now we consider a double strip, consisting of two coupled strips for spinless fermions, see Fig. \ref{fig:3d}. This system is equivalent to a single strip but for spinful fermions.
Below we focus on the double strip but we note that one can identify the upper (lower) strip with spin up (down) state labeled by $\eta=1$ ($\eta=-1$). 
The chemical potentials $\mu_\eta$ are opposite for the two strips, $\mu_1=-\mu_{\bar 1}=\sqrt{2}t_x$, and are chosen such that the system is at half-filling. For the spinful strip the role of 
$\mu_\eta$ is
played by the Zeeman term, $\mu_\eta=\eta g \mu_B B$, arising from the magnetic field $\bf B$  along $z$. Here, $g$ is the $g$-factor, and $\mu_B$ is the Bohr magneton. The inter-strip hopping amplitude $t_z$ is also accompanied by the phase $\phi^z$ arising from a uniform magnetic field ${\bf B}_2$ applied along $y$,
\begin{equation}
H_z = \sum_{n, m, \sigma, \eta} t_z e^{i n \eta \phi^z } c_{n,m,\sigma,\eta}^\dagger c_{n,m,\sigma,\bar{\eta}}.
\end{equation}
The amplitude of  ${\bf B}_2$ is chosen  so that $\phi^z=(e/\hbar c) B_2 a_x a_z=\pi$. 
This amounts to apply a total field ${\bf B}_{tot} = {\bf B}+{\bf B}_2$  in 
the $yz$-plane. Moreover, the same $H_z$ is generated in the spinful case  by a $B_2$-field applied along  $y$ with an amplitude that oscillates in space along $x$ with period $2a_x$, or, alternatively, by Rashba spin orbit interaction. \cite{Braunecker_Loss_Klin_2009}

Again, we search for wavefunctions in terms of right and left mover fields defined around two Fermi points, $k_{F1}=\pi/4a_x$ (upper strip) and $k_{F\bar{1}}=3\pi/4a_x$ (lower strip).
The  linearized Hamiltonian density for this extended model in terms of the Pauli matrices $\eta_i$ acting on the upper/lower strip subspace is given by [see Eq.~(\ref{density})]
\begin{align}
\mathcal{H}_2 = \mathcal{H} (\sigma_2 \to \sigma_2 \eta_3) + t_z \eta_1 \tau_1\,.
\label{density2}
\end{align}
The resulting spectrum is
$\epsilon_{l,\pm,p}=\pm\sqrt{(\hbar \upsilon_F  k)^2 + (t_{yl}+p t_z)^2}$, with $p=\pm1$.
We note that the gap vanishes if $|t_z|=t_{y1}$ or $|t_z|=t_{y2}$. The closing and reopening of a gap often signals a topological phase transition. Indeed, imposing vanishing boundary conditions at the edges, we find that there are two edge states  (one at each edge)  at zero energy, $E=0$, if the following topological criterion is satisfied, $t_{y1}<|t_z|<t_{y2}$, see Fig. \ref{fig:states_3_top}. The wavefunction of the left edge state for $t_z>0$ is 
given by (with $x=na_x$)
\begin{align}
&\psi^L_{E=0}=
(f(x),-if^*(x),i(-1)^nf(x),(-1)^{n+1}f^*(x)),\nonumber\\
&f(x)=e^{-ik_y a_y/2}e^{-(k_{2-}+ik_{F1})x}-\cos (k_y a_y/2)\nonumber\\
&\hspace{10pt}\times e^{-(k_{1-}-ik_{F1})x}+i \sin (k_y a_y/2) e^{-(k_{1+}-ik_{F1})x}.
\end{align}
The basis $( \psi_{1,1}, \psi_{1,\bar 1}, \psi_{\bar 1,1}, \psi_{\bar 1,\bar 1})$ is composed of wavefunctions $\psi_{\eta,\sigma}$ defined at the $\sigma$-unit lattice site of the $\eta$-strip. 
The smallest wavevectors $k_{l,\pm} =|t_{l}\pm t_z|/\hbar\upsilon_F$ determine the localization length of the edge state. 
We note that the probability densities $|\psi_{\eta,\sigma}(x)|^2$ are uniform inside the unit cell.

If $|t_z|<t_{y1}$, there are two edge states at each edge for $E$ inside the gap; see Fig. \ref{fig:states_3_top}. These  states, propagating in $y$, have a momentum $k_y$ determined by $E$. 
This case 
is similar to  one  strip with spinless particles discussed above. 
We note that the edge states found here are the higher-dimensional extensions of the end bound states found in one-dimensional nanowires~\cite{Two_field_Klinovaja_Stano_Loss_2012,footnote1} and ladders~\cite{Klinovaja_Loss_FF_1D}.
For  $|t_z|>t_{y2}$, there are no edge states; see Fig. \ref{fig:states_3_top}.
Finally, the most interesting regime here is $t_{y1}<|t_z|<t_{y2}$, where there is one zero-energy edge state  at each edge, see Fig. \ref{fig:states_3_top}. Such states with flat dispersion are  expected to be strongly affected by interactions.

{\it Conclusions.} We have studied topological regimes of  strips with modulated hopping amplitudes in the presence of magnetic fields. 
We found topological regimes with chiral edge states at filling factors that correspond to integer and fractional QHE regimes. We   showed that double strips
sustain topological phases with mid-gap edge states with flat dispersion.
Optical lattices~\cite{Lewenstein} seem to be promising candidates for implementing directly the anisotropic tight-binding models considered here.

This work is supported by the Swiss NSF, NCCR Nanoscience, and NCCR QSIT.

\appendix

\section{Fourier transformation \label{appendix_1}} We introduce the momentum $k_y$ via Fourier transformation,
\begin{equation}
 c_{n, m,\sigma}  = \frac{1}{\sqrt{N_y}}\sum_{k_y} e^{im k_y a_y} c_{n, k_y, \sigma},
\end{equation}
where $N_y$ is the number of  lattice sites in $y$-direction. By analogy, we introduce a momentum $k_x$,
\begin{equation}
c_{n,k_y,\sigma}  = (1/\sqrt{N_x})\sum_{k_x} e^{i n k_x a_x} c_{k_x k_y,\sigma},
\end{equation}
where $N_x$ is the number of lattice sites along $x$.

\section{Effective Hamiltonian \label{appendix_2}}
Here we derive the spectrum of $ H=H_x+ H_y$ in the continuum limit following Refs. \onlinecite{MF_wavefunction_klinovaja_2012, Two_field_Klinovaja_Stano_Loss_2012}.
The annihilation operator in position space $\Psi(x)$ close to the Fermi level is expressed  in terms of slowly varying right [$R_\sigma (x)$] and left [$L_\sigma (x)$] movers as
\begin{equation}
\Psi(x) =\sum_\sigma [R_\sigma (x)  e^{ik_F x} + L_\sigma  (x) e^{-ik_F x}].
\end{equation}
As a consequence, $H_{x}$ results in the kinetic term,
\begin{align}
H_{x}^{lin} = -i\hbar \upsilon_F \sum _\sigma\int dx\ &[R_\sigma^\dagger(x)\partial_xR_\sigma(x) \nonumber \\
&\hspace{30pt}-L_\sigma^\dagger(x)\partial_xL_\sigma(x)],
\end{align}
and  $ H_{y}$ results in a term  that couples right and left movers, 
\begin{align}
&H_{y}^{lin}=  t_{y1} \int dx\  \big[L_1^\dagger(x)R_{\bar 1}(x) + h.c.]\nonumber\\
&\hspace{30pt} +t_{y2} \int dx\ [e^{-i k_y a_y } R_1^\dagger(x)L_{\bar 1}(x) +  h.c.\big].
\end{align}
Here, we used the specific choice of the parameters, $2k_F a_x=\phi\in (0,\pi)$. 
It is this term that leads to resonant scattering  and opens the gaps at the Fermi points.
The Fermi velocity $\upsilon_F$ depends on the Fermi wavevector, $\hbar \upsilon_F = 2t_x a_x \sin(k_F a_x)$.
The Hamiltonian density $\mathcal{H}$ corresponding to $H^{lin}=H_{x}^{lin}+H_{y}^{lin}=\int dx\ \Psi^\dagger \mathcal{H}\Psi$, can be rewritten in terms of the Pauli matrices $\tau_i$ ($\sigma_i$), acting on the right-left mover (lattice) subspace, $\Psi = (R_1, L_1, R_{\bar{1}}, L_{\bar{1}})$, leading directly to Eq.~(\ref{density}) in the main text.

\section{ Wavefunction of left edge state \label{appendix_3}}
The wavefunction of the state localized at the left edge of a spinless strip is given by
\begin{align}
&\psi_{k_y}(x)= 
\begin{pmatrix}
e^{-k_2 x-i(k_{F}x +\theta)} - e^{-k_1 x+i (k_{F}x-\theta)}  \\
e^{-k_2 x+ik_{F}x} - e^{-k_1 x-i k_{F}x}  
  \end{pmatrix},\label{wavefunction}
\end{align}
where we suppress the normalization factor. Here, we introduced the notations
$e^{i \theta}=(E+i\sqrt{t_{y1}^2-E^2})/t_{y1}$ and
$k_l = \sqrt{t_{yl}^2-E^2}/\hbar \upsilon_F$.


\begin{thebibliography}{100}
\bibitem{footnote_filling} Here we use the standard definition of the filling factor $\nu=N/(BS/(hc/e))$, where $S$ is the strip area.
\bibitem{footnote:B_period} 
At given $\nu$, the  $x$ periodicity of the full stripe model is given by $n \pi/k_F$ (with reciprocal vector $2k_F/n$).
However, in leading order (in $t_y$) only the $K$-periodicity of the unperturped Bloch functions  enters.
We emphasize that, in contrast to the lattice model, the period of our stripe model depends explicitly on $k_F$. This is possible only in the presence of interactions: one additional electron
changes the period for all electrons.
\end{thebibliography}
\end{document}